\documentclass[12pt]{article}

\usepackage{amsthm,amsmath, amssymb, amsfonts, amscd, xspace, pifont}
\usepackage{epsfig, psfrag, psfig, pstricks}
\usepackage{float, fancybox, fullpage}
\usepackage{delarray}
\usepackage{hyperref}
\usepackage{url}
\usepackage{anysize}
\usepackage{multirow}
\usepackage{enumerate}
\usepackage{latexsym,ifthen}
\usepackage{subfigure}
\usepackage{anysize}
\usepackage[authoryear,round]{natbib}

\newtheorem{theorem}{Theorem}

\def\xv{\mathbf x}

\def\1v{\mathbf 1}

\def\0v{\mathbf 0}
\def\bx{\mathbf x}
\def\muv{\boldsymbol \mu}

\begin{document}

\title{Statistical Significance of Clustering using Soft Thresholding}

\author{Hanwen Huang$^1$, Yufeng Liu$^{2,3,4,5}$, Ming Yuan $^6$,
   and J. S. Marron$^{2,3,4}$ \\ 
    {\it $^1$ Department of Epidemiology and Biostatistics}\\ 
    {\it University of Georgia, Athens, GA 30605}\\ 
    {\it $^2$ Department of Statistics and Operations Research}\\ 
    {\it $^3$ Department of Biostatistics} \\ 
    {\it $^4$ Lineberger Comprehensive Cancer Center}\\ 
    {\it $^5$ Carolina Center for Genome Sciences} \\ 
    {\it University of North Carolina at Chapel Hill}\\ 
    {\it Chapel Hill, NC 27599}\\ 
    {\it $^6$ Department of Statistics}\\
    {\it University of Wisconsin-Madison}\\ 
    {\it Madison, WI 53706} \\ 
    email:huanghw@uga.edu\\
          ~yfliu@email.unc.edu\\
          ~~~~~myuan@stat.wisc.edu\\
          ~~~~marron@email.unc.edu\\
          } \date{}

\maketitle

\begin{abstract}

Clustering methods have led to a number of important discoveries in
bioinformatics and beyond. A major challenge in their use is
determining which clusters represent important underlying structure,
as opposed to spurious sampling artifacts. This challenge is
especially serious, and very few methods are available, when the data
are very high in dimension. Statistical Significance of Clustering
(SigClust) is a recently developed cluster evaluation tool for high
dimensional low sample size data. An important component of the
SigClust approach is the very definition of a single cluster as a
subset of data sampled from a multivariate Gaussian distribution. The
implementation of SigClust requires the estimation of the eigenvalues
of the covariance matrix for the null multivariate Gaussian
distribution. We show that the original eigenvalue estimation can lead
to a test that suffers from severe inflation of type-I error, in the
important case where there are a few very large eigenvalues. This
paper addresses this critical challenge using a novel likelihood based
soft thresholding approach to estimate these eigenvalues, which leads
to a much improved SigClust. Major improvements in SigClust
performance are shown by both mathematical analysis, based on the new
notion of Theoretical Cluster Index, and extensive simulation
studies. Applications to some cancer genomic data further demonstrate
the usefulness of these improvements.

\end{abstract}


\section{Introduction}
Clustering methods have been broadly applied in many fields including
biomedical and genetic research. They aim to find data structure by
identifying groups that are similar in some sense. Clustering is a
common step in the exploratory analysis of data. Many clustering
algorithms have been proposed in the literature (see
\cite{Duda-2000-PCL-url,Hastie:elementsStat:2001} for comprehensive
reviews).  Clustering is an important example of unsupervised
learning, in the sense that there are no class labels provided for the
analysis.  Clustering algorithms can give any desired number of
clusters, which on some occasions have yielded important scientific
discoveries, but can also easily be quite spurious. This motivates
some natural cluster evaluation questions such as:
\begin{itemize}
\item how can the statistical significance of a clustering result be
  assessed?
\item are clusters really there or are they mere artifacts of
  sampling fluctuations?
\item how can the correct number of clusters for a given data
  set be estimated?
\end{itemize}

Several cluster evaluation methods have been
developed. \cite{McShaneRFYLS02} proposed a cluster hypothesis test
for microarray data by assuming that important cluster structure in
the data lies in the subspace of the first three principal components,
where standard low dimensional methods can be
used. \cite{citeulike:310156} proposed using resampling techniques to
evaluate the prediction strength of different clusters.
\cite{citeulike:825687} wrote an R package for assessing the
significance of hierarchical clustering. Despite progress in this
area, evaluating significance of clustering remains a serious
challenge, especially in High Dimensional Low Sample Size (HDLSS)
situations.

Numerous works on the application of Gaussian mixture models to
cluster analysis have appeared in the literature. Overviews can be
found in \cite{mclachlan2000,fraley2002}. Gaussian mixture models need
estimation of the full parameters of each component, which can be
quite challenging when tackling HDLSS problems. Recently, some
regularization-based techniques have been applied to model-based
clustering of high-dimensional data, see e.g.
\cite{pan2007,wang2008,xie2008,McNicholas2010,Baek2011}. A more recent
review in this area can be found in \cite{Bouveyron2014}.

\cite{liu:sigclust} proposed a Monte Carlo based method called
Statistical Significance of Clustering (SigClust) which was
specifically designed to assess the significance of clustering results
for HDLSS data. An important contribution of that paper included a
careful examination of the question of ``what is a cluster?". For this
reason, with an eye firmly on the very challenging HDLSS case, their
notion of cluster was taken to be ``data generated from a single
multivariate Gaussian distribution".  This Gaussian definition of
``cluster" has been previously used by \citet{kou1993} and
\citet{mclachlan2000} and \citet{fraley2002}. This was a specific
choice, which made the HDLSS problem tractable, but entailed some
important consequences. For example, none of the Cauchy, Uniform, nor
even $t$ distributions correspond to a single cluster in this
sense. While this may seem to be a strong assumption, it has allowed
sensible real data analysis in otherwise very challenging HDLSS
situations, with a strong record of usefulness in bioinformatics
applications, see e.g. \cite{sigapply1,Neil10}. From this perspective,
SigClust formulates the problem as a hypothesis testing procedure with
\begin{itemize}
\item[] $H_0$: the data are from a single Gaussian distribution
\item[] $H_1$: the data are not from a single Gaussian distribution.
\end{itemize}
As noted in \cite{liu:sigclust}, this choice of null hypothesis is
more sensible than say a difference between subgroups in terms of
means, because clustering methods will split even a truely Gaussian
population into (mean based) statistically significant subgroups. The
test statistic used in SigClust is the 2-means cluster index which is
defined as the ratio of the within cluster variation to the total
variation. Because this statistic is location and rotation invariant,
it is enough to work only with a Gaussian null distribution with mean
$0$ and diagonal covariance matrix $\Lambda$. The null distribution of
the test statistic can be approximated empirically using a direct
Monte Carlo simulation procedure. The significance of a clustering
result can be assessed by computing an appropriate
$p$-value. Recently, \cite{maitra2012} proposed a non-parametric
bootstrap approach for assessing significance in the clustering of
multidimensional datasets. They defined a cluster to be a subset of
data sampled from a spherically symmetric, compact and unimodal
distribution and a non-parametric version of the bootstrap was used to
sample the null distribution. It is important to note that their
method has not been developed to handle HDLSS situations yet.

SigClust has given useful and reasonable answers in many high
dimensional applications
(\cite{milano2008,sigapply1,Neil10}). However, SigClust was based on
some approximations and left room for improvement. In order to
simulate the null distribution of the test statistic, SigClust uses
invariance principles to reduce the problem to just estimating a
diagonal null covariance matrix. This is the same task as finding the
underlying eigenvalues of the covariance matrix. Therefore, a key step
in SigClust is the effective estimation of these
eigenvalues. Currently a factor analysis model is used to reduce the
covariance matrix eigenvalue estimation problem to the problem of
estimating a low rank component that models biological effects
together with a common background noise level. However, the empirical
studies in Section \ref{simulation} show that when there are a few
large eigenvalues this method can be dramatically improved.

Recently, many sparse methods have been introduced to improve the
estimation of the high dimensional covariance matrix, see e.g.
\cite{Meinshausen06highdimensional,yuanlin2007,friedman2008,zhu2008,
fan2009,witten2009,yuan2010,cai2011,witten2011}, among many
others. Let $d$ denote the dimension of the feature space. Then, a
critical difference between SigClust and these sparse approaches is
that SigClust only needs estimates of the $d$ eigenvalues instead of
the $d(d-1)/2$ parameters of the full covariance matrix. This is
because, based on the rotation invariance of the test statistic, the
null distribution of the test statistic used in SigClust is determined
by the eigenvalues rather than the entire covariance
matrix. Nevertheless, the sparse methods are somewhat related to the
soft thresholding method proposed in this paper, in the sense that the
latter also incorporates a type of $L_1$ penalty.

The contributions in this paper start by showing that, when the first
eigenvalue is huge, the original SigClust (which in Section
\ref{likehard} is seen to be reasonably called $hard~thresholded$) can
be seriously anti-conservative. This behavior is studied using the
novel notion of Theoretical Cluster Index. This also motivates an
appropriate soft thresholding variation, which is seen to give vastly
improved SigClust performance over a very wide range of settings,
through both theoretical analysis and detailed simulations. The rest
of the article is organized as follows. In Section \ref{method}, we
first give a brief description of the SigClust procedure and the
existing hard thresholding eigenvalue estimation approach. Then we use
mathematical analysis, together with likelihood ideas to develop the
new soft thresholding approach. To compare the performances of
different methods, numerical studies are given in Section
\ref{simulation} for simulated and in Section \ref{real} for real data
examples.  We provide some discussion in Section \ref{discussion} and
collect proofs of the likelihood derivation in the supplementary
material.

\section{Methodology}\label{method}
In this section, we first briefly review the SigClust method in
Section \ref{review}. In Section \ref{likehard}, we provide an
alternative likelihood based derivation, based on the hard
thresholding ideas, for the estimation of the covariance matrix
eigenvalues used in the original SigClust paper. Then we introduce a
new soft thresholding approach, based on replacing the hard $L_0$
constraint, with a softer $L_1$ constraint, in Section
\ref{soft}. Insightful mathematical analysis, based on the new notion
of Theoretical Cluster Index, which gives deep insights into the
performance of these methods, and also suggests a reasonable tuning
approach, is given in Section \ref{ci}.

\subsection{Review of the Original SigClust Method}\label{review}
SigClust is a clustering evaluation tool which can be used to assess
the significance of a given clustering result by providing an
appropriate $p$-value. Suppose that the original data set $X$, of
dimension $d\times n$, has $d$ variables and $n$ observations,
i.e. $X=[\bx_1,\cdots,\bx_n]$, where each $\bx_i\in R^d$. The null
hypothesis of SigClust is that the data are from a single Gaussian
distribution $N({\bf{\mu}},\Sigma)$, where ${\bf{\mu}}$ is a
$d$-dimensional vector and $\Sigma$ is a $d\times d$ covariance
matrix. Given a clustering of the vectors in $X$, i.e. sets $C_1$ and
$C_2$, where $C_1\cup C_2=\{1,...,n\}$ and $C_1$ and $C_2$ are
disjoint, the strength of the clusters can be assessed using the two
means cluster index (CI), which is the sum of the within class
variation divided by the total variation. More precisely,
\begin{eqnarray}
  \text{CI}=\frac{\sum_{k=1}^2\sum_{i\in C_k}\|\bx_i-\bar{\bx}^{(k)}\|^2}
       {\sum_{i=1}^n\|\bx_i-\bar{\bx}\|^2}, 
\end{eqnarray}
where $\bar{\bx}^{(k)}$ represents the mean of the $k$th cluster for
$k=1,2$ and $\bar{\bx}$ represents the overall mean. CI can be based
on either an input clustering, or can be calculated by a conventional
k-means algorithm \citep{citeulike:4238358}. SigClust uses CI as the
test statistic which has the nice property of being both location and
rotation invariant. This leads to a dramatic reduction in the number
of parameters to be estimated. In particular, during simulation, the
mean $\mu$ can be taken to be ${\bf{0}}$, because of the location
invariance. In a parallel way, rotation invariance provides a major
reduction in the parametrization of $\Sigma$ to a diagonal matrix
$\Lambda=\text{diag}(\lambda_1,\cdots,\lambda_d)$, using the
eigen-decomposition $\Sigma=U\Lambda U^T$, where $U$ is an orthogonal
matrix (essentially a rotation matrix). A factor analysis model is
used to estimate the $d$ eigenvalues which are still a relatively
large number of parameters compared with the sample size $n$ for HDLSS
data sets. Specifically, $\Lambda$ is modeled as
\begin{eqnarray}
  \Lambda=\Lambda_0+\sigma_N^2I,
\end{eqnarray}
where the diagonal matrix $\Lambda_0$ represents true underlying
biology and is typically low-dimensional, and $\sigma_N^2$ represents
the level of background noise. This model is identifiable because many
of the diagonal elements of $\Lambda_0$ are zero. First $\sigma_N$ is
estimated as
\begin{eqnarray}\label{mad}
  \hat{\sigma}_N=\frac{\text{MAD}_{d\times n~\text{data
        set}}}{\text{MAD}_{N(0,1)}},
\end{eqnarray}
where MAD stands for the median absolute deviation from the median,
and $\hat{\sigma}_N$ is rescaled by MAD$_{N(0,1)}=0.6745$, the
theoretical MAD of the N(0,1) distribution, to be on the correct
scale. Then $\Lambda$ is estimated to be
\begin{eqnarray}\label{hard}
  \hat{\lambda}_j=\left\{\begin{array}{lr}
  \tilde{\lambda}_j & \mbox{if}~\tilde{\lambda}_j\ge\hat{\sigma}_N^2\\
  \hat{\sigma}_N^2 & \mbox{if}~\tilde{\lambda}_j<\hat{\sigma}_N^2,
  \end{array}\right.
\end{eqnarray}
where $(\tilde{\lambda}_1,\cdots,\tilde{\lambda}_d)$ are the
eigenvalues of the sample covariance matrix. 

The procedure for SigClust can be briefly summarized as follows: 
\begin{itemize}
\item[] Step 1. Calculate the cluster index for the original data set
  based on the given two-cluster assignments. The cluster assignments
  can be obtained from previous knowledge about the data or from
  application of a clustering algorithm such as k-means.
\item[] Step 2. Obtain estimates
  $(\hat{\lambda}_1,\cdots,\hat{\lambda}_d)$ for the eigenvalues
  $(\lambda_1,\cdots,\lambda_d)$ of $\Sigma$.
\item[] Step 3. Simulate data $N_\text{sim}$ times with each data set
  consisting of $n$ i.i.d. observations from the null distribution
  $(x_1,\cdots,x_d)$ with $x_j$ drawn independently from the
  $N(0,\hat{\lambda}_j)$ distribution. Here $N_\text{sim}$ is some large
  number.
\item[] Step 4. Calculate the corresponding cluster index for each
  simulated data set from Step 3 to obtain an empirical distribution
  of the cluster index based on the null hypothesis.
\item[] Step 5. Calculate a $p$-value (based on an empirical quantile)
  for the original data set and draw a conclusion based on a
  prespecified test level. 
\end{itemize}

\subsection{Likelihood Interpretation of the SigClust Method: Hard 
Thresholding}\label{likehard} Now we first show that the solution
(\ref{hard}) can also be obtained based on a more general likelihood
consideration, which gives the SigClust method of \cite{liu:sigclust}
a new interpretation. In factor models, the covariance matrix can be
written as
\begin{equation}
  \Sigma=\Sigma_0+\sigma_N^2I
\end{equation}
for some low rank positive semi-definite matrix $\Sigma_0$. Denote the
precision matrix
\begin{eqnarray}
  C\equiv\Sigma^{-1}\equiv(\Sigma_0+\sigma_N^2I)^{-1}=\frac{1}{\sigma_N^2}I-W_0
\end{eqnarray}
for some positive semi-definite matrix $W_0$ with
rank$(\Sigma_0)$=rank$(W_0)$.
 
To estimate $\Sigma$, we minimize the negative log-likelihood to yield
the following semi-definite program
\begin{eqnarray}\label{sdp0}
  &\text{argmin}_{C}\Big[-\log|C|+\text{trace}(C\tilde{\Sigma})\Big],
  \quad\\\label{sdp1} \mbox{subject to}\quad &C=\frac{1}{\sigma_N^2}
  I-W_0,~C,W_0\succeq 0,
\end{eqnarray}
where the sample covariance is denoted as
$\tilde{\Sigma}=(1/n)(X-\bar{X})(X-\bar{X})^T$ and $A\succeq 0$ means
that $A$ is positive semi-definite. 

In factor models, we want to encourage a small number of factors which
amounts to encouraging a small rank for $\Sigma_0$ or $W_0$. The
direct approach to enforcing low rank $\Sigma_0$ or $W_0$ is to add an
extra rank constraint:
\begin{eqnarray}\label{con1}
  \text{rank}(W_0)\leq l,
\end{eqnarray}
where $l$ is a pre-specified tuning parameter. Denote the
eigen-decomposition $W_0=UDU^T$ and
$\tilde{\Sigma}=\tilde{U}\tilde{\Lambda}\tilde{U}^T$, where
$D=\text{diag}(d_1,\cdots,d_d)$ and
$\tilde{\Lambda}=\text{diag}(\tilde{\lambda}_1,\cdots,\tilde{\lambda}_d)$.
Then $C=U(\frac{1}{\sigma_N^2}I -D)U^T$.
\begin{theorem}\label{thm1}
  For a fixed $\sigma^2_N$, the solution to
  (\ref{sdp0}),(\ref{sdp1}),(\ref{con1}) is given by $U=\tilde{U}$ and
  \begin{equation}
    \hat{d}_k=\left\{\begin{array}{ll} \frac{1}{\sigma_N^2}
    -\frac{1}{\tilde{\lambda}_k} &
    \text{if} ~ k\leq l~\mbox{and}~\tilde{\lambda}_k >\sigma_N^2\\ 0 &
    \text{otherwise}.
    \end{array}\right.
  \end{equation}
\end{theorem}
Proof of this Theorem and other proofs are given in the supplementary
material. By Theorem \ref{thm1}, we get the eigenvalues of the
covariance matrix which are identical to (\ref{hard}) with suitable
choices of $l$, i.e. greater than or equal to the number of
eigenvalues which are bigger than $\sigma_N^2$ given by
(\ref{hard}). We call this estimation the hard thresholding approach,
so this name applies to the estimation used in \cite{liu:sigclust} as
described in Section \ref{review} above.

\subsection{Soft Thresholding Approach}\label{soft}

\begin{figure}[hbtp] \vspace{0cm} 
    \begin{center}
      \epsfig{file=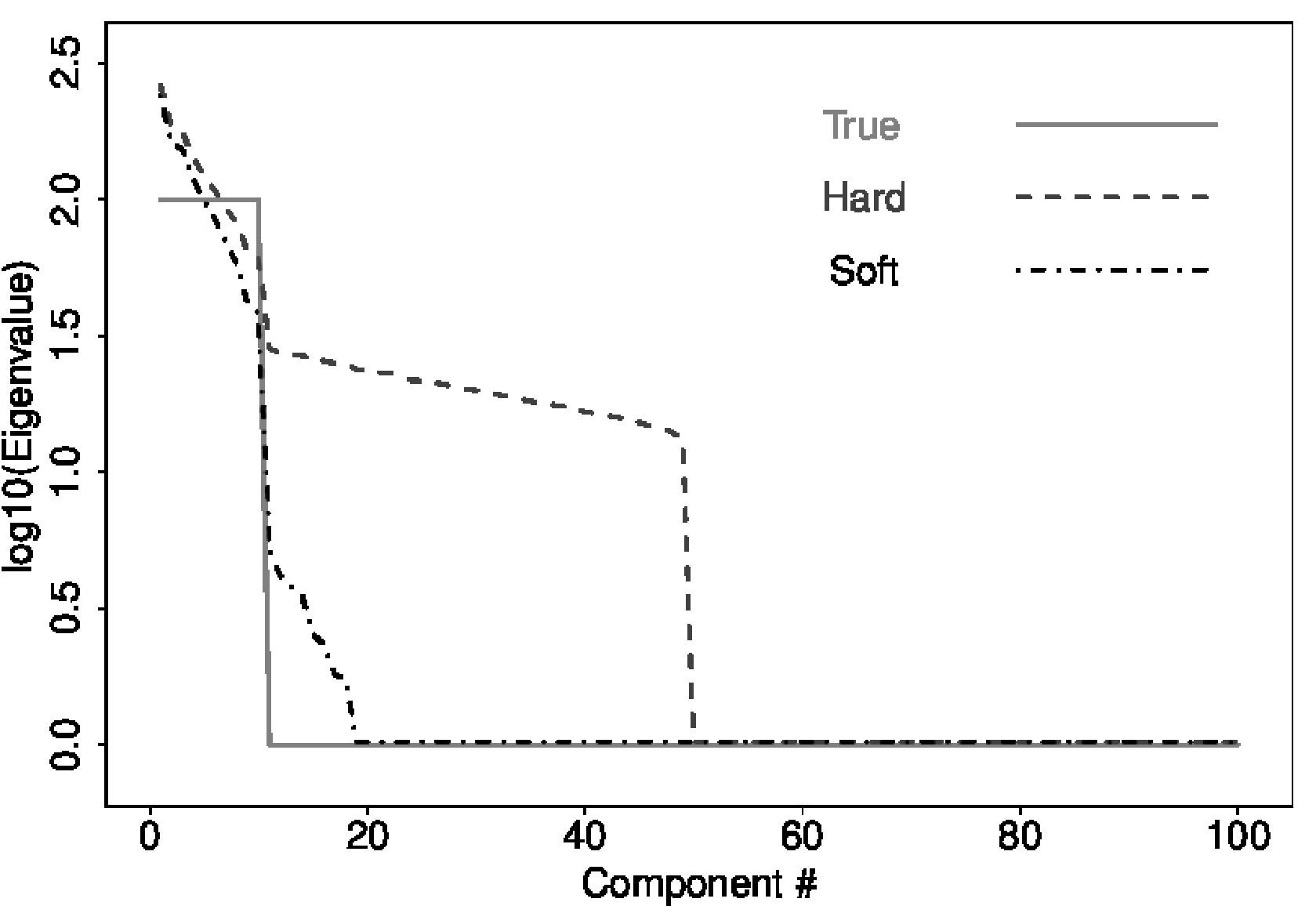,
	width=10cm,totalheight=13cm,angle=-90}
    \end{center} \vspace{-0cm} 
    \caption{True and estimated covariance matrix eigenvalues based on
      the hard- and soft-thresholding methods for a simulated data set
      with $d=1000$ and $n=50$. This shows that some eigenvalues are
      highly over-estimated by the hard thresholding method. The soft
      thresholding method gives major improvement for this example.}
    \vspace{0cm}
    \label{figure1}
\end{figure}

As mentioned in \cite{liu:sigclust}, a challenge for the hard
thresholding method is the effective estimation of the large
eigenvalues in HDLSS settings. This is illustrated in Figure
\ref{figure1} using a simple HDLSS example with $n=50$ and
$d=1000$. The data are generated from a multivariate normal
distribution with covariance matrix $\Lambda$, where $\Lambda$ is
diagonal with elements $(\underbrace{v,\cdots,v}_w,1,\cdots,1)$.  We
consider $v=100$ and $w=10$, which gives the true eigenvalue (solid)
curve in Figure \ref{figure1}. In the HDLSS setting, it is well known
that the sample estimators of the larger eigenvalues tend to grossly
overestimate the corresponding true eigenvalues, because the total
variation (both signal and noise) that is spread over the 1000
dimensional data set is concentrated in the first 50 non-zero
eigenvectors. Note that for the first $50$ entries the hard
thresholding estimates (dashed curve) are the same as the sample
eigenvalues. The rest of the sample eigenvalues are not shown, as they
are $0$, which cannot be plotted on this log scale. For entries
11--50, the estimated eigenvalues are far too high, because they have
been raised by the concentration of the noise energy into these
terms. This effect is precisely quantified in \cite{rmt06}. This
tendency for the hard thresholding to overestimate the true
eigenvalues can create anti-conservatism in SigClust, which will be
mathematically analyzed in Section \ref{ci}. In this section we
propose a less aggressive thresholding scheme which can appropriately
reduce the larger estimated eigenvalues to avoid this source of bias
towards larger eigenvalues, called $soft~thresholding$, shown as the
dot-dashed curve in Figure \ref{figure1}.

Our approach is to replace the rank constraint (\ref{con1}) on $W_0$
by a smooth constraint. In particular, the $L_0$ constraint
(\ref{con1}) becomes an $L_1$ constraint
\begin{eqnarray}\label{con2}
  &\text{trace}(W_0)\le M
\end{eqnarray}
where the signal versus noise trade-off is controlled by a tuning
parameter $M\ge 0$. Denote $\lambda_{max}(W_0)$ the largest eigenvalue
of $W_0$, according to Theorem 1 of \citet{Fazel2001},
trace$(W_0)\le\lambda_{max}(W_0)$rank$(W_0)$. Therefore, the
constraint above is a convex envelope to rank$(W_0)$ (in the sense of
being the biggest convex function below rank($W_0$)) and therefore a
convex relaxation of the constraint on rank$(W_0)$. The tuning value
$M=0$ results in $W_0=0$, since $W_0\succeq 0$, so by (5) this gives a
pure noise model with $\Sigma=\sigma^2I$.  As $M$ increases, the
non-noise component $W_0$ plays a stronger role in the estimation. The
constraint (\ref{con2}) is a nuclear norm constraint, which has been
well-studied in the convex optimization literature, see e.g.
\cite{Fazel2002}.

Solution of the soft thresholding optimization problem, which is
determined by (7), (8) and (11) is given in a closed form in Theorem 2
below.
\begin{theorem}
  For a fixed $\sigma^2_N$, given $M$, the solution to the soft
  thresholding optimization problem:
  \begin{eqnarray}\nonumber
    &\text{argmin}_{C}\Big[-\log|C|+\text{trace}(C\tilde{\Sigma})\Big]
    \quad\\\nonumber \mbox{subject to}\quad &C=\frac{1}{\sigma_N^2}
    I-W_0,~C,W_0\succeq 0,~\text{trace}(W_0)\le M
  \end{eqnarray}
  is given by $U=\tilde{U}$ as in Theorem 1 with soft thresholded
  estimated eigenvalues 
  \begin{eqnarray}\label{softest}
    \hat{\lambda}_k=\left\{\begin{array}{lr} \tilde{\lambda}_k-\tau&
    \text{if}~~\tilde{\lambda}_k>\tau+\sigma_N^2\\ \sigma_N^2 &
    \text{if}~~\tilde{\lambda}_k\le \tau+\sigma_N^2
    \end{array}\right.
    =(\tilde{\lambda}_k-\tau-\sigma_N^2)_++\sigma_N^2,
  \end{eqnarray}
  where $\tau$ is the solution of the equation
  \begin{equation}\label{tau}
    \sum_{k=1}^d\left(\frac{1}{\sigma_N^2}
    -\frac{1}{(\tilde{\lambda}_k-\tau)_+}\right)_+=M.
  \end{equation}
\end{theorem}

Note that the solution of this optimization problem (now using the
constraint (\ref{tau})) is essentially a constant downward shift of
the estimated sample eigenvalues by the value $\tau$, while still
keeping them above $\sigma_N^2$. Because of this intuition and the
relationship (\ref{tau}) between $\tau$ and $M$, it is useful to think
of $\tau$ as a more insightful, but equivalent, version of the tuning
parameter than $M$.

Figure \ref{figure1} shows the improvements available in eigenvalue
estimation, for that HDLSS example, from soft thresholding. The
constant downward shift in estimated eigenvalues is hard to interpret
because of the log scale used there. The consequences of this in terms
of SigClust are demonstrated mathematically in Section 2.4 and through
simulation in Table \ref{table1}. A reasonable basis for choice of the
tuning parameter $\tau$ will be derived from the concept of
$Theoretical~Cluster~Index$ (TCI), defined in Section \ref{ci}.

\subsection{Theoretical Gaussian 2-means Cluster Index}\label{ci}

Once the covariance matrix eigenvalues are estimated, we can proceed
with the SigClust analysis. Toward that end, we need to determine the
null distribution of the 2-means cluster index. In this section, we
will derive a theoretical relationship between the cluster index and
the eigenvalues which clarifies the performance of both hard and soft
thresholding, and also leads to a useful choice of $\tau$ in the
latter case.

As above, let $\xv=(x_1,\cdots,x_d)$ be a $d$-dimensional random
vector having a multivariate normal distribution of $\xv\sim
N(0,\Sigma)$ with mean 0 and covariance matrix $\Sigma=U\Lambda U^T$.
Denote $\phi(.)$ the multivariate normal probability density function
with mean ${\bf 0}$ and variance $I$. Define the theoretical total sum
of squares as
\begin{eqnarray}
  {\text{TSS}}=E\|\xv\|^2=\int\|\xv\|^2\phi(\xv)d\xv.
\end{eqnarray} 
The theoretical within cluster sum of squares (WSS) is based on a
theoretical analog of clusters, which is a partition of the entire
feature space $R^d$ into ${\cal S}_1$ and ${\cal S}_2$. Define
$\muv_1=\int_{\xv\in{\cal S}_1}\xv\phi(\xv)d\xv/\int_{\xv\in{\cal
S}_1}\phi(\xv)d\xv$ and $\muv_2=\int_{\xv\in{\cal
S}_2}\xv\phi(\xv)d\xv/\int_{\xv\in{\cal S}_2}\phi(\xv)d\xv$. Then
we have
\begin{eqnarray}
  {\text{WSS}}=\int_{\xv\in{\cal S}_1}\|\xv-\muv_1\|^2\phi(\xv)d\xv+\int_{\xv\in{\cal
      S}_2}\|\xv-\muv_2\|^2\phi(\xv)d\xv.
\end{eqnarray}
These are combined to give TCI=WSS/TSS. The relationship between TCI
and the covariance matrix eigenvalues is stated by the following
theorem.
\begin{theorem}\label{thm3}
  For an optimal choice of ${\cal S}_1$ and ${\cal S}_2$, i.e. the
  split is chosen to minimize the total WSS over all possible splits
  (this is the theoretical analog of 2-means clustering), the
  Theoretical Cluster Index is
  \begin{eqnarray}\label{tci}
    {\text{TCI}}=1-\frac{2}{\pi}\frac{\lambda_1}{\sum_{i=1}^{d}\lambda_i}.
  \end{eqnarray}
\end{theorem}

Theorem \ref{thm3} tells us that the optimal TCI is only determined by
two quantities, the largest eigenvalue $\lambda_1$ and the total sum
of eigenvalues $\sum_{i=1}^d\lambda_i$. In practice, different
eigenvalue estimation methods give quite different estimates of these
two quantities, which in turn leads to quite different SigClust
performances.

For the sample covariance estimation method, the estimated $\lambda_1$
is typically larger (i.e. biased upwards) than the true value, in
HDLSS situations. Since the sum of the sample eigenvalues is a
consistent estimate of the total variation, i.e. the denominator of
(\ref{tci}), it follows that the resulting estimate of TCI will
generally be smaller (biased downwards) than the true TCI in that
case, giving a larger $p$-value and thus a conservative result.

For the hard thresholding method, it follows from (\ref{tci}) that the
sample and hard methods have the same numerator but the hard
thresholding method gives a larger denominator. Thus the hard
thresholding method has larger TCI, and consequently is always more
powerful than the sample method. However, the hard thresholding method
has large potential for creating type-I error. Consider potential
biases in the estimation of $\lambda_1$ and $\sum_{i=1}^d\lambda_i$
defined as $\delta_1$ and $\Delta$ respectively. Some algebra shows
that the essential difference between the true TCI and the hard
thresholding estimate is proportional to
\begin{eqnarray}\label{dhard}
  E=\frac{\lambda_1+\delta_1}{\sum_{i=1}^d\lambda_i+\Delta}
  -\frac{\lambda_1}{\sum_{i=1}^d\lambda_i}=\frac{\sum_{i=1}^d
    \lambda_i\delta_1-\lambda_1\Delta}{\sum_{i=1}^d\lambda_i
    (\sum_{i=1}^d\lambda_i+\Delta)}.
\end{eqnarray}
When 
\begin{eqnarray}\label{19}
  \delta_{1}<\frac{\lambda_1\Delta}{\sum_{i=1}^d\lambda_i}, 
\end{eqnarray}
i.e. when the first eigenvalue is large relative to the rest, the hard
thresholding method will tend to be anti-conservative.

\begin{figure}[hbtp] \vspace{0cm} 
    \begin{center}
      \epsfig{file=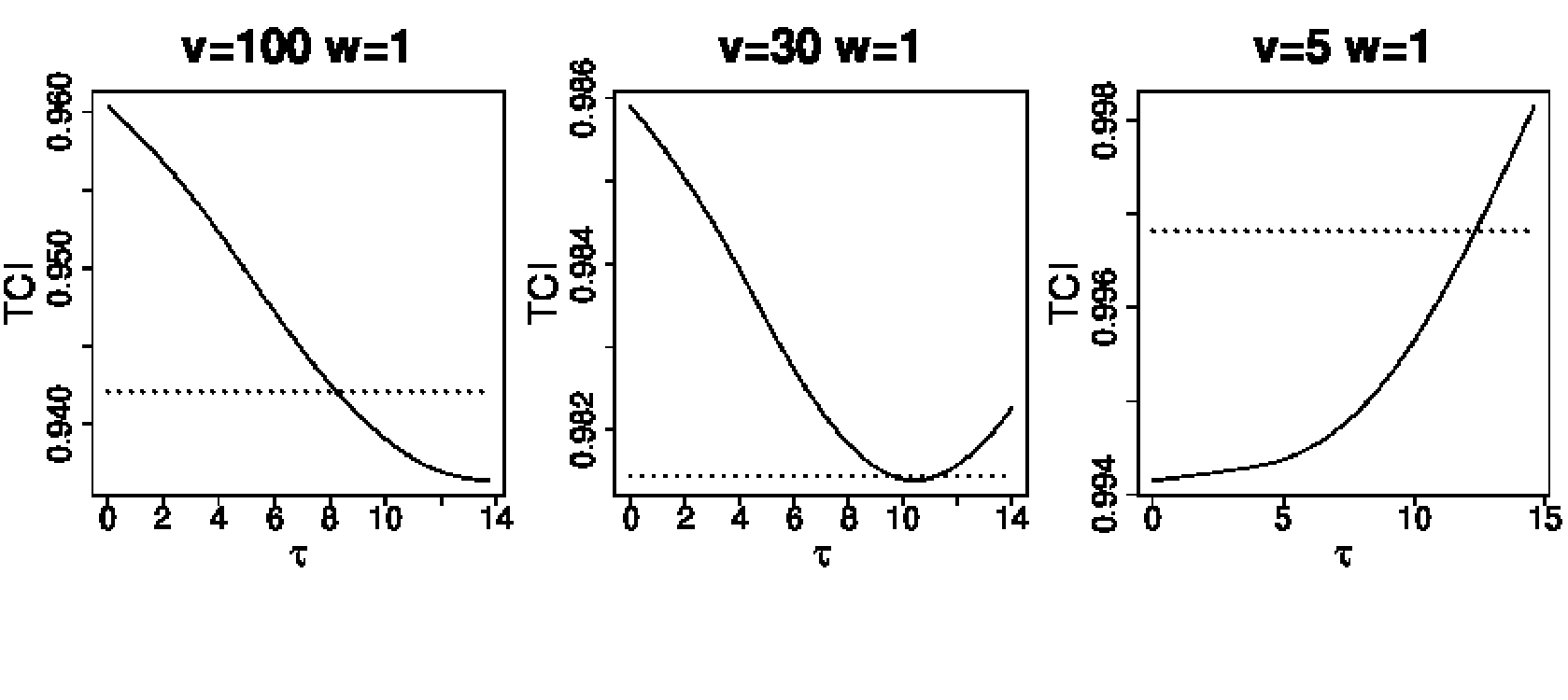,width=6.5cm,totalheight=13.5cm,angle=-90}
    \end{center} \vspace{-1.5cm} 
    \caption{Relationships between TCI and tuning parameter $\tau$ for
    three different settings (solid line). Dotted lines represent the
    TCI calculated from true eigenvalues. Shows situations where hard
    thresholding is anti-conservative (left panel), where energy
    preserving soft thresholding is anti-conservative (right panel),
    and where $\tau^\star$ is strictly between the two (central
    panel).}
    \vspace{-0.5cm}
    \label{tau.figure}
\end{figure}

For the soft thresholding method, the large eigenvalues are subtracted
by $\tau$ from the corresponding sample estimates. This will decrease
both the numerator and denominator of (\ref{tci}) in contrast to the
hard thresholding method. Our goal is to choose $\tau$ so that the
type-I error can be controlled and at the same time the test is more
powerful than the sample method. For this, a useful boundary is
$\tilde{\tau}$, which is energy conserving in the sense that the sum
of the soft eigenvalues is the sum of the sample eigenvalues,
i.e. $\tilde{\tau}$ is the solution of the equation
\begin{eqnarray}\label{14}
  \sum_{k=1}^d\{(\tilde{\lambda}_k-\tilde{\tau}-\sigma_N^2)_+
  +\sigma_N^2\}=\sum_{k=1}^d\tilde{\lambda}_k.
\end{eqnarray}
Another important endpoint of reasonable $\tau$ values, is $\tau=0$,
which corresponds to hard thresholding (i.e. no reduction in
eigenvalues). Note that the soft estimated version of the denominator
of the fraction in (\ref{tci}) (i.e. the sum of the eigenvalues) is
monotone decreasing in $\tau$. At $\tilde{\tau}$, this denominator is
the same as the sample version. It follows that for $\tau$ between 0
and $\tilde\tau$, soft thresholding gives bigger TCI than the sample
covariance estimate, and thus a more powerful test. Figure 2 displays
the relationship between TCI calculated from the eigenvalues estimated
using the soft method and tuning parameter $\tau$ for three simulated
data sets with $d=1000$, $n=100$, $w=1$ and $v=100,30,5$ respectively.
In Figure 2, the range of $\tau$ is $[0,\tilde\tau]$. Depending on the
context, SigClust can be anti-conservative at either end of the
interval $[0,\tilde{\tau}]$. This will happen at $\tau=0$, when the
first eigenvalue is very large relative to the others (as shown in the
left panel of Figure \ref{tau.figure}), and at $\tau=\tilde{\tau}$
when the first is only a little larger than the background noise (as
happens in the right panel of Figure \ref{tau.figure}). The central
panel of Figure \ref{tau.figure} shows a compromise situation, where
both endpoints can be anti-conservative, but there is a more
reasonable choice in between. Because anti-conservatism can happen at
either end of this interval, we recommend choosing the soft
thresholded $\tau$ as conservative as possible, by taking $\tau^\star$
in $[0,\tilde\tau]$ to minimize TCI. The examples in Figure 2 show
that, depending on the setting, $\tau^\star$ can be either 0 (hard
thresholding, left), $\tilde\tau$ (energy preserving, right) or
something in between (central panel).

\section{Simulation}\label{simulation}

In this section we investigate how the estimation of the covariance
matrix eigenvalue affects the SigClust performance using extensive
simulation studies. Four SigClust $p$-value computation methods are
compared using the true covariance matrix as well as estimates from
the sample, hard and soft thresholding approaches, which are referred
to using those names.

We have performed simulations in both low and high dimensional
situations. Here we focus on high dimensional results, because our
main contribution is in HDLSS settings. Three types of examples are
generated here including situations under both null and alternative
hypotheses. The sample size is $n=100$, dimension is $d=1000$, and the
number of data sets generated for each run is $N_\text{sim}=1000$.
Empirical quantiles, as discussed in Section 2.1, were used in each
case. We evaluate different methods based on the criterion of whether
or not they can maximize the power while controlling the type-I
error. In Section \ref{onecluster}, we consider examples of data under
the null hypothesis, i.e., having only one cluster generated by a
single Gaussian distribution. In each example we check the type-I
error of each version of SigClust, for a wide variety of Gaussian null
distributions, by studying how often it incorrectly rejects the null
hypothesis $H_0$. In Sections \ref{twoone} and \ref{twoall}, we
explore the power of the different versions of SigClust, by
considering data from a collection of mixtures of two Gaussian
distributions with different signal sizes counting how often it
correctly rejects the null hypothesis. We give combined discussion of
the simulation results in Section \ref{summary}.

\subsection{Level of the Test}\label{onecluster}

In order to evaluate the Type I error rates for different methods,
data were generated under the null hypothesis, i.e. from a single
multivariate Gaussian distribution with $d=1000$ dimensional
covariance matrix $\Lambda$ which is diagonal with elements
$(\underbrace{v,\cdots,v}_w,1,\cdots,1)$. We consider 31 combinations
of $v$ and $w$ with $v=1,\cdots,1000$, and the corresponding
$w=1,\cdots,100$, as shown in Table 1. The simulation procedure was
repeated 100 times for each setting.

Table 1 summarizes the mean and the number of the $p$-values which are
less than 0.05 (N5) and 0.1 (N10) based on different methods under the
various parameter settings. Theoretically the $p$-value follows the
uniform $[0,1]$ distribution since the data are generated from a
single Gaussian distribution. As expected, the empirical distributions
of $p$-values using the true method are relatively close to the
uniform distribution. The sample method results in $p$-values whose
means are always bigger than the expected ones. This is consistent
with the theoretical results shown in Section \ref{ci}. The N5 and N10
are almost all $0$, so we conclude that the sample method is
conservative in all of these settings. For settings of $v\ge 30$,
i.e. for populations with a generally large first few eigenvalues
(e.g. a strongly elongated distribution), with the exception of
$(v,w)=(40,25)$, results based on the hard thresholding method
exhibits more small $p$-values than expected under the uniform
distribution which implies that this approach is anti-conservative in
that situation. On the other hand, the hard thresholding method tends
to be quite conservative, for relatively small values of $v$, e.g. for
approximately more spherical Gaussian distributions. This also is a
consequence of equation (\ref{dhard}), because $\lambda_1=v$ is small.

For the soft method, the results of Table \ref{table1} show
conservative results in all settings.  Like the sample method, the
soft method effectively controls type-I error under the null
hypothesis. But more importantly it can dramatically increase the
power over the sample method under important alternative hypotheses as
shown in the next section. 

The means of the $p$-value populations give additional insights, and
the results are mostly consistent with those from the quantiles. In
particular, the means of the $p$-value from SigClust using the true
eigenvalues are generally close to the desired value of 0.5, the means
from the sample method tend to be larger, and the hard, soft fluctuate
in a way that corresponds to their quantile behavior. An important
point is that means from the soft method are generally substantially
closer to 0.5 than is true for either the sample or the hard methods.

\begin{table}\footnotesize
  \begin{center}
    \caption{Summary table of empirical SigClust $p$-value distribution
    over 100 replications based on four methods under different
    settings in Simulation \ref{onecluster}. The mean and the numbers
    of $p$-values which are less than 0.05 (denoted as N5) and 0.1
    (denoted as N10) are reported ($d=1000$, $n=100$).}\label{table1}
    \begin{tabular}{rrrrrrrrrrrrrr}
    \vspace{-0mm}
      &&\multicolumn{3}{c}{True}&\multicolumn{3}{c}{Sample}&\multicolumn{3}{c}{Hard}
    &\multicolumn{3}{c}{Soft}\\ 
      $v$ & $w$ & Mean & N5 & N10 & Mean & N5 & N10 & Mean & N5 & N10 & Mean & N5 & N10\\ 
      \hline
      1000 & 1 & 0.47 & 5 & 8 & 0.52 & 0 & 1 & 0.00 & 100 & 100 & 0.46 & 0 & 2 \\ 
      200 & 5 & 0.39 & 4 & 10 & 0.82 & 0 & 0 & 0.01 & 94 & 100 & 0.69 & 0 & 0 \\ 
      100 & 10 & 0.34 & 7 & 14 & 0.94 & 0 & 0 & 0.08 & 39 & 75 & 0.82 & 0 & 0 \\ 
      40 & 25 & 0.31 & 7 & 15 & 1.00 & 0 & 0 & 0.56 & 0 & 0 & 0.96 & 0 & 0 \\ 
      20 & 50 & 0.26 & 7 & 17 & 1.00 & 0 & 0 & 0.96 & 0 & 0 & 1.00 & 0 & 0 \\ 
      10 & 100 & 0.21 & 21 & 32 & 1.00 & 0 & 0 & 1.00 & 0 & 0 & 1.00 & 0 & 0 \\ 
      200 & 1 & 0.49 & 6 & 11 & 0.61 & 0 & 0 & 0.00 & 100 & 100 & 0.40 & 0 & 0 \\ 
      100 & 1 & 0.47 & 2 & 7 & 0.78 & 0 & 0 & 0.00 & 100 & 100 & 0.41 & 0 & 1 \\ 
      50 & 1 & 0.49 & 3 & 5 & 0.93 & 0 & 0 & 0.00 & 99 & 100 & 0.34 & 0 & 1 \\ 
      40 & 1 & 0.52 & 7 & 9 & 0.96 & 0 & 0 & 0.01 & 91 & 98 & 0.33 & 0 & 0 \\ 
      30 & 1 & 0.54 & 4 & 11 & 0.99 & 0 & 0 & 0.07 & 54 & 78 & 0.35 & 0 & 2 \\ 
      20 & 1 & 0.53 & 6 & 13 & 1.00 & 0 & 0 & 0.48 & 0 & 9 & 0.49 & 0 & 0 \\ 
      10 & 1 & 0.47 & 4 & 10 & 1.00 & 0 & 0 & 1.00 & 0 & 0 & 0.97 & 0 & 0 \\ 
      50 & 10 & 0.41 & 9 & 15 & 0.98 & 0 & 0 & 0.07 & 48 & 77 & 0.79 & 0 & 0 \\ 
      40 & 10 & 0.35 & 9 & 13 & 0.99 & 0 & 0 & 0.07 & 53 & 77 & 0.74 & 0 & 0 \\ 
      30 & 10 & 0.37 & 11 & 15 & 1.00 & 0 & 0 & 0.12 & 22 & 50 & 0.73 & 0 & 0 \\ 
      20 & 10 & 0.35 & 8 & 15 & 1.00 & 0 & 0 & 0.36 & 2 & 6 & 0.73 & 0 & 0 \\ 
      10 & 10 & 0.35 & 8 & 15 & 1.00 & 0 & 0 & 0.98 & 0 & 0 & 0.93 & 0 & 0 \\ 
      50 & 5 & 0.35 & 7 & 21 & 0.96 & 0 & 0 & 0.01 & 99 & 100 & 0.58 & 0 & 0 \\ 
      40 & 5 & 0.39 & 4 & 13 & 0.98 & 0 & 0 & 0.01 & 94 & 100 & 0.58 & 0 & 0 \\ 
      30 & 5 & 0.37 & 10 & 19 & 0.99 & 0 & 0 & 0.04 & 64 & 91 & 0.54 & 0 & 0 \\ 
      20 & 5 & 0.41 & 5 & 12 & 1.00 & 0 & 0 & 0.30 & 3 & 15 & 0.59 & 0 & 0 \\ 
      10 & 5 & 0.35 & 8 & 14 & 1.00 & 0 & 0 & 0.98 & 0 & 0 & 0.92 & 0 & 0 \\ 
      50 & 2 & 0.43 & 5 & 9 & 0.94 & 0 & 0 & 0.00 & 100 & 100 & 0.43 & 0 & 0 \\ 
      40 & 2 & 0.43 & 10 & 20 & 0.97 & 0 & 0 & 0.01 & 98 & 99 & 0.41 & 0 & 0 \\ 
      30 & 2 & 0.49 & 5 & 9 & 0.99 & 0 & 0 & 0.05 & 66 & 85 & 0.41 & 0 & 1 \\ 
      20 & 2 & 0.45 & 5 & 11 & 1.00 & 0 & 0 & 0.35 & 3 & 12 & 0.49 & 0 & 1 \\ 
      10 & 2 & 0.43 & 8 & 17 & 1.00 & 0 & 0 & 1.00 & 0 & 0 & 0.95 & 0 & 0 \\ 
      5 & 1 & 0.20 & 22 & 38 & 1.00 & 0 & 0 & 1.00 & 0 & 0 & 1.00 & 0 & 0 \\ 
      3 & 1 & 0.16 & 24 & 45 & 1.00 & 0 & 0 & 1.00 & 0 & 0 & 1.00 & 0 & 0 \\ 
      1 & 1 & 0.16 & 19 & 40 & 1.00 & 0 & 0 & 1.00 & 0 & 0 & 1.00 & 0 & 0 \\ 
      \hline
    \end{tabular}
  \end{center}
\end{table}

\subsection{Power of Test for Signal in One Coordinate Direction}\label{twoone}

\begin{figure}[hbtp] \vspace{0cm} 
    \begin{center}
      \epsfig{file=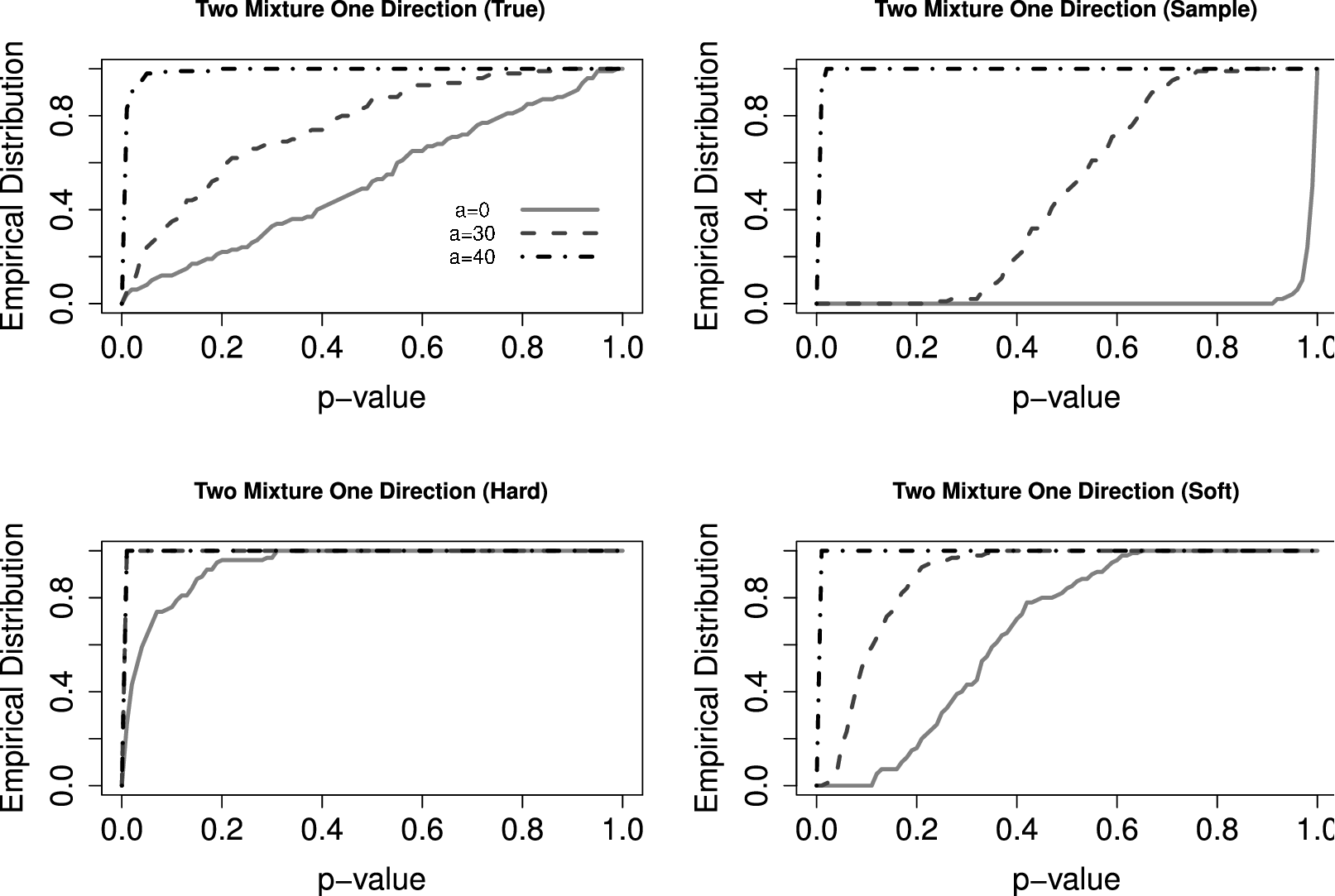,height=15.5cm,angle=-90}
    \end{center} \vspace{-0cm} 
    \caption{Empirical distributions of SigClust $p$-values for
      Simulation \ref{twoone}. This shows sample is too conservative,
      hard is anti-conservative, while soft strikes a nice balance in
      overall performance.}
    \vspace{0cm}
    \label{figure2}
\end{figure}

In this section, we compare the power properties of these various
SigClust hypothesis tests. This is based on a mean mixture of two
normal distributions, $.5N(0,\Lambda)+.5N(\mu,\Lambda)$, where
$\mu=(a,0,\cdots,0)$ with $a=0,30,40$ and as above
$\Lambda=\text{diag}(\underbrace{v,\cdots,v}_w,1,\cdots,1)$ a diagonal
matrix. Here we focus on a similar setting as in the middle panel of
Figure 2 with $v=30$ and $w=1$. When $a=0$, the distribution reduces
to a single Gaussian distribution. The larger the $a$, the stronger
the signal. The theoretical null distribution is $N(0,\Lambda^*)$,
where
$\Lambda^*=\text{diag}(\lambda_1+0.25a^2,\lambda_2,\cdots,\lambda_d)$. The
empirical distributions of $p$-values based on 100 replications are
shown in Figure \ref{figure2}. As expected, the true method (upper
left) is very powerful under the alternative hypothesis and meanwhile
can control the type-I error well under the null hypothesis
($a=0$). The hard method (lower left) is grossly anti-conservative,
since the solid curve bends far above the diagonal. The sample method
(upper right) is too conservative, since the solid curve bends way
below, and even for $a=30$, is very low. The soft method (lower right)
is close to the diagonal at $a=0$, and is more powerful than the
sample, in terms of bending upwards when there is signal in the data.

\subsection{Power of Test for Signal in All 
Coordinate Directions}\label{twoall}

\begin{figure}[hbtp] \vspace{0cm} 
    \begin{center}
      \epsfig{file=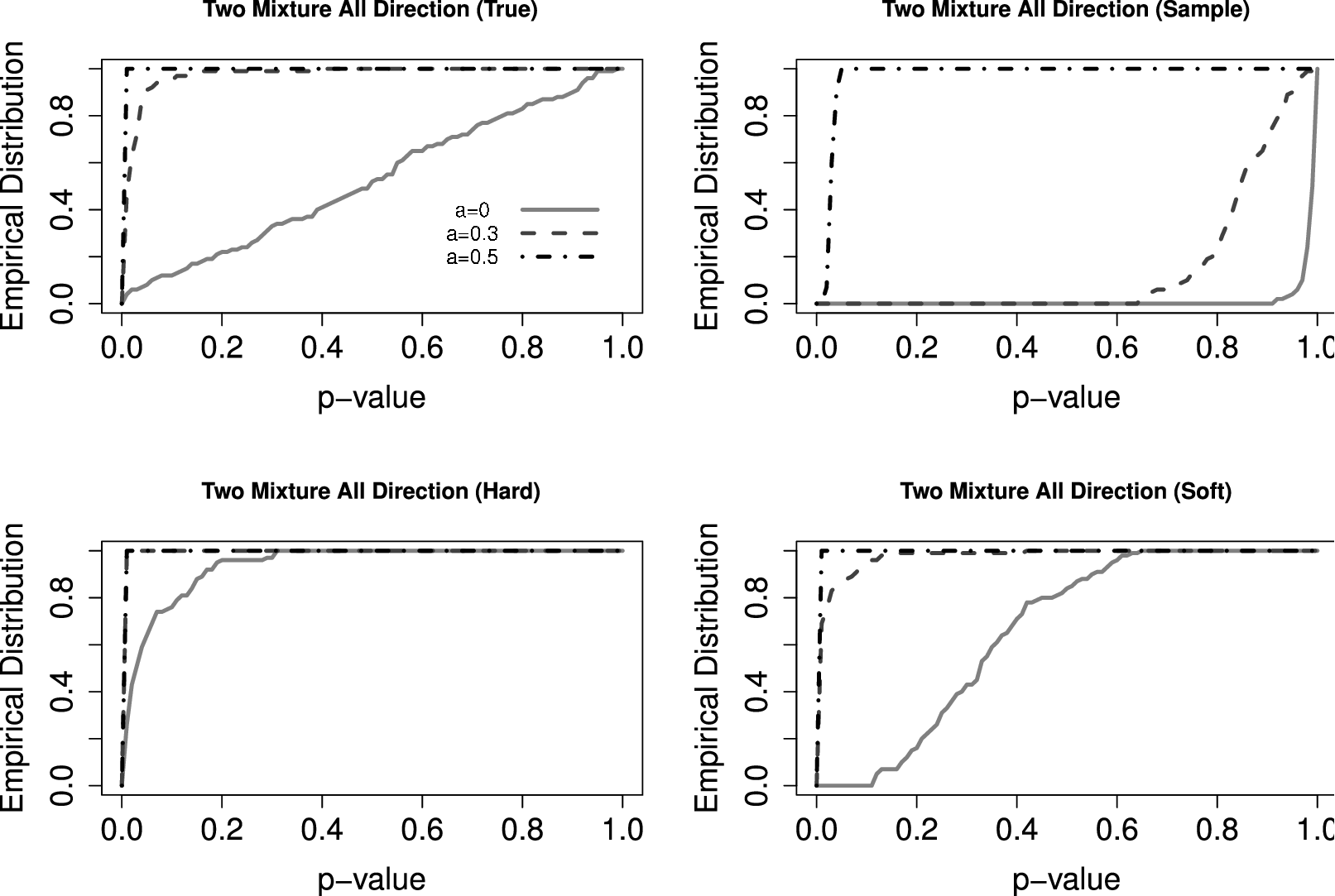,totalheight=15.5cm,angle=-90}
    \end{center} \vspace{-0cm} 
    \caption{Empirical distributions of SigClust $p$-values for
      Simulation \ref{twoall}. The results indicate that hard is
      strongly anti-conservative, while sample is too conservative.
      Overall best is the soft method.}
    \vspace{0cm}
    \label{figure3}
\end{figure}

In the previous subsection, the signal is only in the first coordinate
direction. Now we consider power using another example with the signal
in all coordinate directions.  Similarly, we generate data from a
mixture of two Gaussian distributions,
$.5N(0,\Lambda)+.5N(\muv,\Lambda)$, where $\muv=(a,a,\cdots,a)$ with
$a=0,0.3,0.5$ and
$\Lambda=\text{diag}(\underbrace{v,\cdots,v}_w,1,\cdots,1)$ with
$v=30$ and $w=1$ which is similar to the setting in the previous
section. This signal is very small in each direction, but can be large
when all directions are combined together. The empirical distributions
of $p$-values calculated from the 100 simulated datasets based on
different methods are displayed in Figure \ref{figure3}. For $a=0$ the
results are identical to the single cluster situation in Sections
\ref{onecluster} and \ref{twoone} with $(v,w)=(30,1)$. The hard
thresholding method always yields smaller $p$-values than expected and
thus is strongly anti-conservative under the null hypothesis meaning
it may not be trusted. In contrast, the soft method is conservative
under the null but becomes powerful as the signal increases. When the
signal is big enough, e.g.  $a=0.5$, all methods can identify the
significant clusters. For small signal situations, e.g. $a=0.3$, the
soft method is much more powerful than the sample method.

\subsection{Simulation Summary}\label{summary}
In summary, the sample method is strongly conservative and the hard
thresholding method can be anti-conservative in certain situations. The
soft thresholding method is in-between. Simulation results shown in
Section \ref{onecluster} suggest that, under the null hypothesis, the
performances of the hard thresholding method vary from strongly
conservative to strongly anti-conservative depending on the situations
which are mainly characterized by the quantity, $v$.  The soft method
yields conservative results in all settings studied here. Simulation
results from Sections \ref{twoone} and \ref{twoall} suggest that,
under the alternative hypothesis, the hard thresholding method often
has the largest power and the sample method has the smallest
power. The soft method is appropriately in-between. If the signals are
large enough, all methods can identify the significant
clusters. However, in situations with relatively small signal, the
sample method cannot distinguish the significant clusters. In
practice, we recommend the soft method, i.e., small $p$-values from
the soft method reliably indicate the existence of distinct clusters.

\section{Real Data}\label{real}

In this section, we apply our methods to some real cancer data sets.
As mentioned in \cite{Neil10}, GlioBlastoma Multiforme (GBM) is one of
the most common forms of malignant brain cancer in adults. For the
purposes of the current analysis, we considered a cohort of patients
from The Cancer Genome Atlas Research Network \citep{tcga2010} with
GBM cancer whose brain samples were assayed. 
Four
clinically relevant subtypes were identified using integrated genomic
analysis in \cite{Neil10}, they are Proneural, Neural, Classical, and
Mesenchymal. As in \cite{liu:sigclust}, we filter the genes using the
ratio of the sample standard deviation and sample mean of each
gene. After gene filtering, the data set contained 383 patients with
2727 genes. Among the 383 samples, there are 117 Mesenchymal samples,
69 Neural samples, 96 Proneural samples, and 101 Classical samples.

We applied SigClust to every possible pair-wise combination of
subclasses and calculated the $p$-value based on the three different
methods. Here the cluster-index is computed based on the given class
label. Let MES stand for Mesenchymal samples and CL stand for
Classical samples. Except the MES and CL pair, the $p$-values from all
three methods were significant for each of the five other pairs which
is consistent with the widely accepted fact that these are distinct
clusters. For the MES and CL pair, the $p$-value was non-significant
using the sample method (0.93) while it was significant both for hard
($<0.001$) and soft ($0.04$) methods. This suggests that the sample
was too conservative to find this important effect.  Furthermore the
hard method appears to give too strong a significance, consistent with
its occasional theoretically predicted anti-conservatism.

The second real example we considered is a breast cancer data set
(BRCA) also from The Cancer Genome Atlas Research Network which
include four subtypes: LumA, LumB, Her2 and Basal and have been
extensively studied by microarray and hierarchical clustering analysis
\citep{cfan2006}. The sample size is 343 and the number of genes used
in the analysis after filtering is 4000. Among 343 samples, there are
154 LumA, 81 LumB, 42 Her2 and 66 Basal. 

\begin{figure}[hbtp] \vspace{0cm} 
    \begin{center}
      \epsfig{file=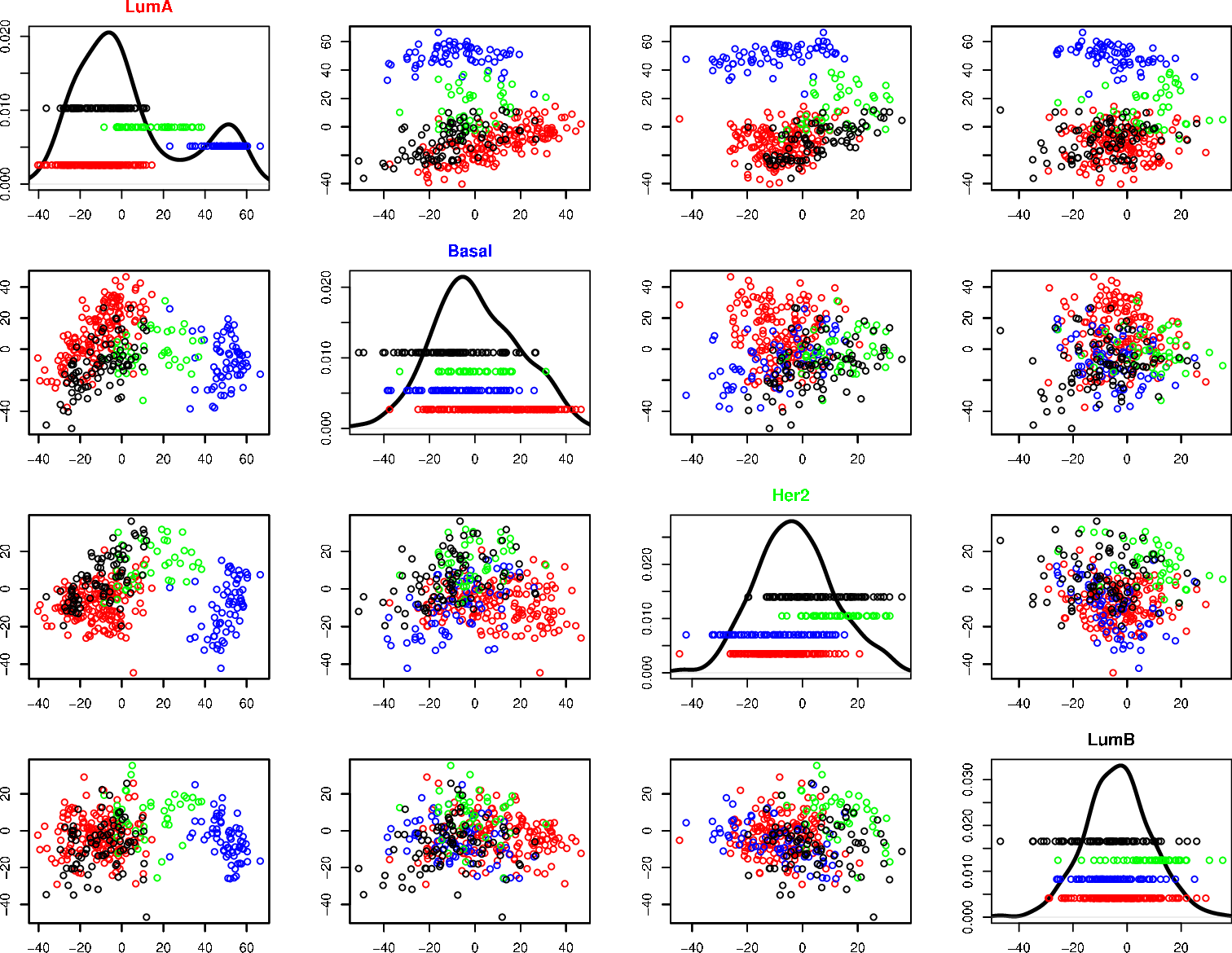,width=13cm,totalheight=14.5cm,angle=-90}
    \end{center} \vspace{-0cm} 
    \caption{PCA projection scatter plot view of the BRCA data,
      showing 1D (diagonal) and 2D projections of the data onto PC
      directions. Groupings of colors and symbols indicate biological
      subtypes. Shows Basals are quite distinct from the others, and
      there is no strong evidence showing that LumA and LumB do not
      come from a single Gaussian distribution.}
    \vspace{0cm}
    \label{figure4}
\end{figure}

The results of applying SigClust to each pair of subclasses are shown
in Table 2. For pairs including Basal, the $p$-values from all three
methods are significant which implies that the Basal cluster is well
separated from the rest. For the LumA and LumB pair, all methods
report very high $p$-values, which suggests that they are actually one
subtype. This is consistent with the findings of \citet{Parker2009},
which suggest that these are essentially a stretched Gaussian
distribution (thus not flagged by SigClust), with an important
clinical division within that distribution. For the Her2 and LumB
pair, all three methods give a non-significant $p$-value although not
as big as for the LumA and LumB pair, so there is no strong evidence
for them to be separated (although hard thresholding appears to be
close to a spuriously significant result). For the Her2 and LumA pair,
the hard method gives a very significant $p$-value, the soft method
gives a nearly significant $p$-value, whereas the sample method fails
to find the clusters. Thus the cluster difference for this pair is not
as large as for the Her2 and LumB pair. Note that the $p$-values
listed in Table \ref{table2} are consistent with the scatter plot in
Figure \ref{figure4} where the projections of the data points onto the
first four principal component (PC) directions are displayed. Clearly,
Basal is well separated from the remaining data. LumA and LumB are
close together and LumB and Her2 are closer than LumA and Her2. These
results are again consistent with our theoretical results that sample
is often conservative, hard can occasionally be anti-conservative, and
soft is an appropriate balance of these effects.

\begin{table}\footnotesize
  \begin{center}
    \caption{SigClust $p$-values for each pair of subtypes for the
      BRCA data. The known class labels are used to calculate the
      cluster index. }
    \label{table2}
    \begin{tabular}{crrrrrr}\\\hline
      &Basal.LumA&Basal.LumB&Basal.Her2&LumA.LumB&Her2.LumB&Her2.LumA\\\hline
      Sample&$<0.001$&$<0.001$&0.015&1&0.99&0.77\\
      Hard&$<0.001$&$<0.001$&$<0.001$&0.89&0.051&$<0.001$\\
      Soft&$<0.001$&$<0.001$&$<0.001$&1&0.95&0.059\\\hline
    \end{tabular}
  \end{center}
\end{table}

\section{Discussion}\label{discussion}

In this paper, we developed a soft thresholding approach and examined
its application to the SigClust method. We found that both the newly
proposed soft thresholding and the hard thresholding proposed in the
original SigClust paper can be derived under a likelihood based
framework and usefully compared in terms of penalties in their
respective regularizations ($L_0$ regularization for hard and $L_1$
for soft). Differences in performance were analyzed using the notion
of Theoretical Cluster Index, which also indicated how the soft method
should be tuned. Through extensive simulation, we compared the
performance of the SigClust method based on different approaches in a
wide variety of settings. As theoretically predicted, we found that
the hard thresholding method would sometimes incorrectly reject the
null while the sample and soft methods are always conservative. The
soft approach was seen to have much better power properties than using
the simple sample covariance estimation. We recommend that our newly
proposed soft method be used in practice because it has been shown to
control the type-I error as well as the sample method under the null
hypothesis, while gaining much more power under the alternative
hypothesis.

An important point is that our definition of $clusters$ assumes a
null Gaussian distribution. Thus SigClust may have limited use in
categorical situations. However, in very high dimensional contexts,
it may be possible to develop a limiting distribution theory by which
most SigClust ideas still work.  Another interesting open problem is
finding a non-simulation calculation of the SigClust p-values.

This paper treats the case of 2 clusters, using a statistic based on
k-means clustering. Interesting extensions include more than 2
clusters, and other clustering criteria, such as those underlying
hierarchical clustering. The incorporation of alternate eigenvalue
estimation methods is another interesting open problem.

In terms of software, the R package for the current version of
SigClust can be freely downloaded on the CRAN website:
http://CRAN.R-project.org/package=sigclust.
Computation time depends on the number of simulated replications, and
the size of the input data. In all cases here, we used
$N_\text{sim}=1000$ and a computer with RAM 16GB and processor 3.5GHz,
and it took around 1 minute for each simulated data set described in
Section 3 and 10 minutes for both the GBM data and the BRCA data
described in Section 4.

\section*{Acknowledgement}
The authors thank the editor, the associate editor, and three referees
for many helpful comments and suggestions which led to a much improved
presentation.

\bibliographystyle{chicago} \bibliography{biblist}

\begin{thebibliography}{}

\bibitem[\protect\citeauthoryear{Baek and McLachlan}{Baek and
  McLachlan}{2011}]{Baek2011}
Baek, J. and G.~J. McLachlan (2011).
\newblock Mixtures of commont-factor analyzers for clustering high-dimensional
  microarray data.
\newblock {\em Bioinformatics\/}~{\em 27}, 1269--1276.

\bibitem[\protect\citeauthoryear{Baik and Silverstein}{Baik and
  Silverstein}{2006}]{rmt06}
Baik, J. and J.~W. Silverstein (2006).
\newblock Eigenvalues of large sample covariance matrices of spiked population
  models.
\newblock {\em Journal of Multivariate Analysis\/}~{\em 97}, 1382--1408.

\bibitem[\protect\citeauthoryear{Bouveyron and Brunet-Saumard}{Bouveyron and
  Brunet-Saumard}{2014}]{Bouveyron2014}
Bouveyron, C. and C.~Brunet-Saumard (2014).
\newblock Model-based clustering of high-dimensional data: A review.
\newblock {\em Computational Statistics and Data Analysis\/}~{\em 71}, 52--78.

\bibitem[\protect\citeauthoryear{Cai, Liu, and Luo}{Cai et~al.}{2011}]{cai2011}
Cai, T.~T., W.~Liu, and X.~Luo (2011).
\newblock A constrained ${L}_1$ minimization approach to sparse precision
  matrix estimation.
\newblock {\em Journal of the American Statistical Association\/}~{\em 106},
  594--607.

\bibitem[\protect\citeauthoryear{Chandriani, Frengen, Cowling, Pendergrass,
  Perou, Whitfield, and Cole}{Chandriani et~al.}{2009}]{sigapply1}
Chandriani, S., E.~Frengen, V.~H. Cowling, S.~A. Pendergrass, C.~M. Perou,
  M.~L. Whitfield, and M.~D. Cole (2009).
\newblock A core {MYC} gene expression signature is prominent in basal-like
  breast cancer but only partially overlaps the core serum response.
\newblock {\em PLoS ONE\/}~{\em 4\/}(8), e6693.

\bibitem[\protect\citeauthoryear{Danaher, Wang, and Witten}{Danaher
  et~al.}{2014}]{witten2011}
Danaher, P., P.~Wang, and D.~Witten (2014).
\newblock The joint graphical lasso for inverse covariance estimation across
  multiple classes.
\newblock To appear in Journal of the Royal Statistical Society, Series B.

\bibitem[\protect\citeauthoryear{Duda, Hart, and Stork}{Duda
  et~al.}{2000}]{Duda-2000-PCL-url}
Duda, R.~O., P.~E. Hart, and D.~G. Stork (2000).
\newblock {\em Pattern Classification}.
\newblock Wiley-Interscience Publication.

\bibitem[\protect\citeauthoryear{Fan, Oh, Wessels, Weigelt, Nuyten, Nobel,
  van't Veer, and Perou}{Fan et~al.}{2006}]{cfan2006}
Fan, C., D.~S. Oh, L.~Wessels, B.~Weigelt, D.~S. Nuyten, A.~B. Nobel, L.~J.
  van't Veer, and C.~M. Perou (2006).
\newblock Concordance among gene-expressionbased predictors for breast cancer.
\newblock {\em New England Journal of Medicine\/}~{\em 355\/}(6), 560--569.

\bibitem[\protect\citeauthoryear{Fan, Feng, and Wu}{Fan et~al.}{2009}]{fan2009}
Fan, J., Y.~Feng, and Y.~Wu (2009).
\newblock Network exploration via the adaptive lasso and {SCAD} penalties.
\newblock {\em The Annals of Applied Statistics\/}~{\em 3}, 521--541.

\bibitem[\protect\citeauthoryear{Fazel}{Fazel}{2002}]{Fazel2002}
Fazel, M. (2002).
\newblock Matrix rank minimization and applications.
\newblock Ph.D. Thesis, Stanford University.

\bibitem[\protect\citeauthoryear{Fazel, Hindi, and Boyd}{Fazel
  et~al.}{2001}]{Fazel2001}
Fazel, M., H.~Hindi, and S.~P. Boyd (2001).
\newblock A rank minimization heuristic with application to minimum order
  system approximation.
\newblock Proceedings of the American Control Conference, 6, 4734-4739.

\bibitem[\protect\citeauthoryear{Fraley and Raftery}{Fraley and
  Raftery}{2002}]{fraley2002}
Fraley, C. and A.~Raftery (2002).
\newblock Model-based clustering, discriminant analysis, and density
  estimation.
\newblock {\em Journal of the American Statistical Association\/}~{\em 97},
  611--631.

\bibitem[\protect\citeauthoryear{Friedman, Hastie, and Tibshirani}{Friedman
  et~al.}{2008}]{friedman2008}
Friedman, J.~H., T.~Hastie, and R.~Tibshirani (2008).
\newblock Sparse inverse covariance estimation with the graphical lasso.
\newblock {\em Biostatistics\/}~{\em 9}, 432--441.

\bibitem[\protect\citeauthoryear{Hastie, Tibshirani, and Friedman}{Hastie
  et~al.}{2009}]{Hastie:elementsStat:2001}
Hastie, T., R.~Tibshirani, and J.~Friedman (2009).
\newblock {\em The Elements of Statistical Learning\/} (second ed.).
\newblock Springer.

\bibitem[\protect\citeauthoryear{Liu, Hayes, Nobel, and Marron}{Liu
  et~al.}{2008}]{liu:sigclust}
Liu, Y., D.~N. Hayes, A.~Nobel, and J.~S. Marron (2008).
\newblock Statistical significance of clustering for high-dimension, low-sample
  size data.
\newblock {\em Journal of the American Statistical Association\/}~{\em
  103\/}(483), 1281--1293.

\bibitem[\protect\citeauthoryear{MacQueen}{MacQueen}{1967}]{citeulike:4238358}
MacQueen, J. (1967).
\newblock Some methods for classification and analysis of multivariate
  observations.
\newblock In {\em Fifth Berkeley Symposium on Mathematical Statistics and
  Probability}, pp.\  281--297.

\bibitem[\protect\citeauthoryear{Maitra, Melnykov, and Lahiri}{Maitra
  et~al.}{2012}]{maitra2012}
Maitra, R., V.~Melnykov, and S.~N. Lahiri (2012).
\newblock Bootstrapping for significance of compact clusters in
  multidimensional datasets.
\newblock {\em Journal of the American Statistical Association\/}~{\em 107},
  378--392.

\bibitem[\protect\citeauthoryear{Mclachlan and Peel}{Mclachlan and
  Peel}{2000}]{mclachlan2000}
Mclachlan, G. and D.~Peel (2000).
\newblock {\em Finite Mixture Models}.
\newblock New York: Wiley.

\bibitem[\protect\citeauthoryear{McNicholas and Murphy}{McNicholas and
  Murphy}{2010}]{McNicholas2010}
McNicholas, P. and T.~Murphy (2010).
\newblock Model-based clustering of microarray expression data via latent
  gaussian mixture models.
\newblock {\em Bioinformatics\/}~{\em 26}, 2705--2712.

\bibitem[\protect\citeauthoryear{McShane, Radmacher, Freidlin, Yu, Li, and
  Simon}{McShane et~al.}{2002}]{McShaneRFYLS02}
McShane, L.~M., M.~D. Radmacher, B.~Freidlin, R.~Yu, M.-C. Li, and R.~Simon
  (2002).
\newblock Methods for assessing reproducibility of clustering patterns observed
  in analyses of microarray data.
\newblock {\em Bioinformatics\/}~{\em 18\/}(11), 1462--1469.

\bibitem[\protect\citeauthoryear{Meinshausen and B\"uhlmann}{Meinshausen and
  B\"uhlmann}{2006}]{Meinshausen06highdimensional}
Meinshausen, N. and P.~B\"uhlmann (2006).
\newblock High dimensional graphs and variable selection with the lasso.
\newblock {\em Annals of Statistics\/}~{\em 34}, 1436--1462.

\bibitem[\protect\citeauthoryear{Milano, Pendergrass, Sargent, George,
  McCalmont, Connolly, and Whitfield}{Milano et~al.}{2008}]{milano2008}
Milano, A., S.~A. Pendergrass, J.~L. Sargent, L.~K. George, T.~H. McCalmont,
  M.~K. Connolly, and M.~L. Whitfield (2008).
\newblock Molecular subsets in the gene expression signatures of {S}cleroderma
  skin.
\newblock {\em PLoS ONE\/}~{\em 3\/}(7), e2696.

\bibitem[\protect\citeauthoryear{Pan and Shen}{Pan and Shen}{2007}]{pan2007}
Pan, W. and X.~Shen (2007).
\newblock Penalized model-based clustering with application to variable
  selection.
\newblock {\em Journal of Machine Learning Research\/}~{\em 8}, 1145--1164.

\bibitem[\protect\citeauthoryear{Parker, Mullins, Cheang, Leung, Voduc,
  Vickery, Davies, Fauron, He, Hu, Quackenbush, Stijleman, Palazzo, Marron,
  Nobel, Mardis, Nielsen, Ellis, Perou, and Bernard}{Parker
  et~al.}{2009}]{Parker2009}
Parker, J.~S., M.~Mullins, M.~C.~U. Cheang, S.~Leung, D.~Voduc, T.~Vickery,
  S.~Davies, C.~Fauron, X.~He, Z.~Hu, J.~F. Quackenbush, I.~J. Stijleman,
  J.~Palazzo, J.~S. Marron, A.~B. Nobel, E.~Mardis, T.~O. Nielsen, M.~J. Ellis,
  C.~M. Perou, and P.~S. Bernard (2009).
\newblock Supervised risk predictor of breast cancer based on intrinsic
  subtypes.
\newblock {\em Journal of Clinical Oncology\/}~{\em 27\/}(8), 1160--1167.

\bibitem[\protect\citeauthoryear{Rothman, Levina, and Zhu}{Rothman
  et~al.}{2008}]{zhu2008}
Rothman, A., E.~Levina, and J.~Zhu (2008).
\newblock Sparse permutation invariant covariance estimation.
\newblock {\em Electronic Journal of Statistics\/}~{\em 2}, 494--515.

\bibitem[\protect\citeauthoryear{Sarle and Kuo}{Sarle and Kuo}{1993}]{kou1993}
Sarle, W.~S. and A.~H. Kuo (1993).
\newblock The modeclus procedure.
\newblock Technical Report P-256, Cary, NC: SAS Institute Inc.

\bibitem[\protect\citeauthoryear{Suzuki and Shimodaira}{Suzuki and
  Shimodaira}{2006}]{citeulike:825687}
Suzuki, R. and H.~Shimodaira (2006).
\newblock Pvclust: an {R} package for assessing the uncertainty in hierarchical
  clustering.
\newblock {\em Bioinformatics\/}~{\em 22\/}(12), 1540--1542.

\bibitem[\protect\citeauthoryear{TCGA}{TCGA}{2010}]{tcga2010}
TCGA (2010).
\newblock The cancer genome atlas research network.
\newblock
  http://cancergenome.nih.gov/wwd/pilot\_program/research\_network/cgcc.asp.

\bibitem[\protect\citeauthoryear{Tibshirani and Walther}{Tibshirani and
  Walther}{2005}]{citeulike:310156}
Tibshirani, R. and G.~Walther (2005).
\newblock Cluster validation by prediction strength.
\newblock {\em Journal of Computational and Graphical Statistics\/}~{\em
  14\/}(3), 511--528.

\bibitem[\protect\citeauthoryear{Verhaak, Hoadley, Purdom, Wang, Qi, Wilkerson,
  Miller, Ding, Golub, Mesirov, Alexe, Lawrence, O'Kelly, Tamayo, Weir,
  Gabriel, Winckler, Gupta, Jakkula, Feiler, Hodgson, James, Sarkaria, Brennan,
  Kahn, Spellman, Wilson, Speed, Gray, Meyerson, Getz, Perou, Hayes, and
  {Cancer Genome Atlas Research Network}}{Verhaak et~al.}{2010}]{Neil10}
Verhaak, R.~G., K.~A. Hoadley, E.~Purdom, V.~Wang, Y.~Qi, M.~D. Wilkerson,
  C.~R. Miller, L.~Ding, T.~Golub, J.~P. Mesirov, G.~Alexe, M.~Lawrence,
  M.~O'Kelly, P.~Tamayo, B.~A. Weir, S.~Gabriel, W.~Winckler, S.~Gupta,
  L.~Jakkula, H.~S. Feiler, J.~G. Hodgson, C.~D. James, J.~N. Sarkaria,
  C.~Brennan, A.~Kahn, P.~T. Spellman, R.~K. Wilson, T.~P. Speed, J.~W. Gray,
  M.~Meyerson, G.~Getz, C.~M. Perou, D.~N. Hayes, and {Cancer Genome Atlas
  Research Network} (2010).
\newblock Integrated genomic analysis identifies clinically relevant subtypes
  of glioblastoma characterized by abnormalities in {PDGFRA}, {IDH1}, {EGFR},
  and {NF}1.
\newblock {\em Cancer cell\/}~{\em 17\/}(1), 98--110.

\bibitem[\protect\citeauthoryear{Wang and Zhu}{Wang and Zhu}{2008}]{wang2008}
Wang, S. and J.~Zhu (2008).
\newblock Variable selection for model-based high-dimensional clustering and
  its application to microarray data.
\newblock {\em Biometrics\/}~{\em 64}, 440--448.

\bibitem[\protect\citeauthoryear{Witten, Tibshirani, and Hastie}{Witten
  et~al.}{2009}]{witten2009}
Witten, D., R.~Tibshirani, and T.~Hastie (2009).
\newblock A penalized matrix decomposition, with applications to sparse
  principal components and canonical correlation analysis.
\newblock {\em Biostatistics\/}~{\em 10\/}(3), 515--534.

\bibitem[\protect\citeauthoryear{Xie, Pan, and Shen}{Xie
  et~al.}{2008}]{xie2008}
Xie, B., W.~Pan, and X.~Shen (2008).
\newblock Penalized model-based clustering with cluster-specific diagonal
  covariance matrices and grouped variables.
\newblock {\em Electronic Journal of Statistics\/}~{\em 2}, 168--212.

\bibitem[\protect\citeauthoryear{Yuan}{Yuan}{2010}]{yuan2010}
Yuan, M. (2010).
\newblock Sparse inverse covariance matrix estimation via linear programming.
\newblock {\em Journal of Machine Learning Research\/}~{\em 11}, 2261--2286.

\bibitem[\protect\citeauthoryear{Yuan and Lin}{Yuan and
  Lin}{2007}]{yuanlin2007}
Yuan, M. and Y.~Lin (2007).
\newblock Model selection and estimation in the {G}aussian graphical model.
\newblock {\em Biometrika\/}~{\em 94\/}(1), 19--35.

\end{thebibliography}

\end{document}